\documentclass[hidelinks,journal,onecolumn]{IEEEtran}
\usepackage{amsmath,amsfonts,amssymb}
\usepackage{algorithmic}
\usepackage{algorithm}
\usepackage{array}
\usepackage[caption=false,font=footnotesize]{subfig}
\usepackage{textcomp}
\usepackage{stfloats}
\usepackage{url}
\usepackage{verbatim}
\usepackage{graphicx}
\usepackage{cite}
\usepackage{orcidlink}
\usepackage{lipsum}
\usepackage{outlines}
\usepackage{soul}
\usepackage[acronym,shortcuts]{glossaries}
\usepackage[bottom]{footmisc}

\newcommand{\herm}{^{\mathsf{H}}}

\newacronym{ISAC}{ISAC}{integrated sensing and communications}
\newacronym{JCAS}{JCAS}{joint communications and sensing}
\newacronym{OFDM}{OFDM}{orthogonal frequency division multiplexing}
\newacronym{AFDM}{AFDM}{affine frequency division multiplexing}
\newacronym{OCDM}{OCDM}{orthogonal chirp division multiplexing}
\newacronym{OTFS}{OTFS}{orthognal time frequency space}
\newacronym{B5G}{B5G}{beyond fifth generation}
\newacronym{6G}{6G}{sixth generation}
\newacronym{TV}{TV}{time variant}
\newacronym{TI}{TI}{time invariant}
\newacronym{2D}{2D}{two-dimensional}
\newacronym{NTN}{NTN}{non-terrestrial network}
\newacronym{LEO}{LEO}{low earth orbit}
\newacronym{IoT}{IoT}{internet-of-things}
\newacronym{mmWave}{mmWave}{millimeter-wave}
\newacronym{THz}{THz}{Terahertz}
\newacronym{V2X}{V2X}{vehicle-to-everything}
\newacronym{EHF}{EHF}{extremely high-frequency}
\newacronym{ICI}{ICI}{inter-carrier interference}
\newacronym{SotA}{SotA}{state-of-the-art}
\newacronym{UAV}{UAV}{unmanned aerial vehicle}
\newacronym{ZT}{ZT}{Zac transform}
\newacronym{DZT}{DZT}{discrete Zac transform}
\newacronym{IDZT}{IDZT}{inverse discrete Zac transform}
\newacronym{FT}{FT}{Fourier transform}
\newacronym{DFT}{DFT}{discrete Fourier transform}
\newacronym{IDFT}{IDFT}{inverse discrete Fourier transform}
\newacronym{AFT}{AFT}{affine Fourier transform}
\newacronym{DAFT}{DAFT}{discrete affine Fourier transform}
\newacronym{IDAFT}{IDAFT}{inverse discrete affine Fourier transform}
\newacronym{SFFT}{SFFT}{symplectic finite Fourier transform}
\newacronym{ISFFT}{ISFFT}{inverse symplectic finite Fourier transform}
\newacronym{HT}{HT}{Heisenberg transform}
\newacronym{WT}{WT}{Wigner transform}
\newacronym{MIMO}{MIMO}{multiple-input multiple-output}
\newacronym{UE}{UE}{user equipment}
\newacronym{AP}{AP}{access point}
\newacronym{CIR}{CIR}{channel impulse response}
\newacronym{CSI}{CSI}{channel state information}
\newacronym{FMCW}{FMCW}{frequency modulated continuous wave}
\newacronym{AWGN}{AWGN}{additive white Gaussian noise}
\newacronym{ML}{ML}{maximum likelihood}

\begin{document}
\title{ 
    {$~$ \\[-3ex] \large \textit{ \color{red} \textbf{PLEASE FIND THE FULL EXTENDED ARTICLE OF THIS WHITE PAPER HERE: \cite{Rou_SPM24} \\[0.5ex] 
    {\small (ACCEPTED FOR PUBLICATION AT THE IEEE SIGNAL PROCESSING MAGAZINE - SPECIAL ISSUE ON \\[-4ex]  ``SIGNAL PROCESSING FOR THE INTEGRATED SENSING AND COMMUNICATIONS REVOLUTION")}}}}\\
    \!\!\! AFDM vs OTFS: A Comparative Study of Promising \!\!\!\\[-0.25ex] \!\!\! Waveforms for ISAC in Doubly-Dispersive Channels \!\!\!\\[-0.75ex]}

\author{\normalsize 
Hyeon Seok Rou$^{1}$,
Giuseppe Thadeu Freitas de Abreu$^{1}$,
Junil Choi$^{2}$, \\
David Gonz{\'a}lez G.$^{3}$,
Osvaldo Gonsa$^{3}$,
Yong Lian Guan$^{4}$,
Marios Kountouris$^5$.\\[1ex]

\small $^1$School of Computer Science and Engineering, Constructor University, 28759 Bremen, Germany \\[0.1ex]

$^2$School of Electrical Engineering, Korea Advanced Institute of Science and Technology, Daejeon, South Korea \\[0.1ex]

$^{3}$Wireless Communications Technologies, Continental AG, 65936 Frankfurt/Main, Germany \\[0.1ex]

$^{4}$School of Electrical and Electronic Engineering, Nanyang Technological University, 639798 Singapore \\[0.3ex]

$^{5}$Communication Systems Department, EURECOM, Sophia-Antipolis, France\\[1ex]
}

%\markboth{WHITE PAPER Proposal for "title of the special Issue" SPM Special}

\maketitle

%%%%%%%%%%%%%%%%%%%%%%%%%%%%%%%%%%%%%%%%%%%%%%%%%%%%%%%%%%%%%%%%%%%%%%%%%%%%%
%%%%%%%%%%%%%%%%%%%%%%%%%%%%%%%%%%%%%%%%%%%%%%%%%%%%%%%%%%%%%%%%%%%%%%%%%%%%%
\vspace{-7ex}
\section{\textbf{Author Biography}}
\par 
%%%%%%%
\textbf{Hyeon Seok Rou} \textit{(Graduate Student Member, IEEE)} [hrou@constructor.university]  is a Ph.D. Candidate at Constructor University, Bremen, Germany, funded as a Research Associate at Continental AG on a researach project on 6G vehicular-to-everything (V2X) integrated sensing and communications (ISAC).
His research interests lie in the fields of ISAC, hyper-dimensional sparse modulation schemes, B5G/6G V2X wireless communications technology, and Bayesian inference. \\[-2ex]

%%%%%%%
\textbf{Giuseppe Thadeu Freitas de Abreu} \textit{(Senior Member, IEEE)} [gabreu@constructor.university] is a Full Professor of Electrical Engineering at Constructor University, Bremen, Germany. His research interests include communications theory, estimation theory, statistical modeling, wireless localization, cognitive radio, wireless security, MIMO systems, ultrawideband and millimeter wave communications, full-duplex and cognitive radio, compressive sensing, energy harvesting networks, random networks, connected vehicles networks, and many other topics. He has served as an editor for various IEEE Transactions, and currently serves as an editor to the IEEE Signal Processing Letters and the IEEE Communications Letters. \\[-2ex]

%%%%%%%%
\textbf{Junil Choi} \textit{(Senior Member, IEEE)} [junil@kaist.ac.kr]
is a (Named) Ewon Associate Professor with the School of Electrical Engineering, Korea Advanced Institute of Science and Technology (KAIST), South Korea.
From 2007 to 2011, he was a member of the Technical Staff with the Samsung Advanced Institute of Technology (SAIT) and Samsung Electronics Company Ltd., South Korea, where he contributed to advanced codebook and feedback framework designs for the 3GPP LTE/LTE-Advanced and IEEE 802.16m standards. 
His research interests include the design and analysis of massive MIMO, mmWave communication systems, distributed reception, and vehicular communication systems.
He was the recipient of numerous awards from various IEEE societies, and is currently an Associate Editor of IEEE Transactions on Wireless Communications, IEEE Transactions on Communications, IEEE Communications Letters, and IEEE Open Journal of the Communications Society. \\[-2ex]

%%%%%%%%
\textbf{David González G.} \textit{(Senior Member, IEEE)} [david.gonzalez.g@ieee.org] %
is a Senior Research Engineer at Continental AG, Germany, and has previously served at Panasonic Research and Development Center, Germany.
His research interests include aspects of cellular networks and wireless communications, including interference management, radio access modeling and optimization, resource allocation, and vehicular communications. 
Since 2017, he has represented his last two companies as delegate in the 3GPP for 5G standardization, mainly focused on physical layer aspects and vehicular communications. \\[-2ex]

%%%%%%%
\textbf{Osvaldo Gonsa} [osvaldo.gonsa@continental-corporation.com] is the Head of the Wireless Communications Technologies group by Continental AG in Frankfurt, Germany. 
He has worked in research and standardization in radio access network, serving as an advisor to the German Federal Ministry of Economy and Energy for the ``PAiCE" projects, and currently as the GSMA Advisory Board for automotive and the 6GKom project of the German Federal Ministry of Education and Research. \\[-2ex]

%%%%%%%
\textbf{Yong Lian Guan} \textit{(Senior Member, IEEE)} [eylguan@ntu.edu.sg] is a Professor of Communication Engineering with the School of Electrical and Electronic Engineering, Nanyang Technological University (NTU), Singapore, where he leads the Continental-NTU Corporate Laboratory and the successful deployment of the campus-wide NTU-NXP V2X test bed. His research interests broadly include coding and signal processing for communication systems and data storage systems. He is a Distinguished Lecturer of the IEEE Vehicular Technology Society from 2021 to 2023 and an Editor of the IEEE Transactions on Vehicular Technology. \\[-2ex]

%%%%%%%
\textbf{Marios Kountouris} \textit{(Fellow, IEEE)} [kountour@eurecom.fr] is a Professor with the Communication Systems Department, EURECOM, Sophia-Antipolis, France. His research interests include communications theory, machine learning for wireless communications, low latency networking, and stochastic modeling and performance analysis. He was a recipient of the Consolidator Grant of the European Research Council (ERC) in 2020 on goal-oriented semantic communication. He is an AAIA Fellow. He has received several awards and distinctions at various IEEE venues. He has served as an Editor for IEEE Transactions on Wireless Communications, IEEE Transactions on Signal Processing, and IEEE Wireless Communication Letters. He is a Chartered Professional Engineer of the Technical Chamber
of Greece.

%%%%%%%%%%%%%%%%%%%%%%%%%%%%%%%%%%%%%%%%%%%%%%%%%%%%%%%%%%%%%%%%%%%%%%%%%%%%%
%%%%%%%%%%%%%%%%%%%%%%%%%%%%%%%%%%%%%%%%%%%%%%%%%%%%%%%%%%%%%%%%%%%%%%%%%%%%%
\newpage
\section{\textbf{History, motivation, and significance of the topic}}
\label{secrel}
\par 

\textbf{\textit{Abstract:} This white paper aims to briefly describe a proposed article that will provide a thorough comparative study of waveforms designed to exploit the features of doubly-dispersive channels arising in heterogeneous high-mobility scenarios as expected in the \ac{B5G} and \ac{6G}, in relation to their suitability to \ac{ISAC} systems.
In particular, the full article will compare the well-established delay-Doppler domain-based \ac{OTFS} and the recently proposed chirp domain-based \ac{AFDM} waveforms.
Both these waveforms are designed based on a full delay-Doppler representation of the \ac{TV} multipath channel, yielding not only robustness and orthogonality of information symbols in high-mobility scenarios, but also a beneficial implication for environment target detection through the inherent capability of estimating the path delay and Doppler shifts, which are standard radar parameters.
These modulation schemes are distinct candidates for \ac{ISAC} in \ac{B5G}/\ac{6G} systems, such that a thorough study of their advantages, shortcomings, implications to signal processing, and performance of communication and sensing functions are well in order.
In light of the above, a sample of the intended contribution (Special Issue paper) is provided below.}

\vspace{2.5ex}

\Ac{B5G} and \ac{6G} wireless systems rely on \ac{EHF} technology developed for the \ac{mmWave} and \ac{THz} bands \cite{Rappaport_Access19,Chen_CC19,Lopez_WC19}, to provide a variety of enhanced applications such as \Ac{IoT}, edge computing, and smart cities \cite{Vo_MNA22, Saad_Network20}, with particular interest in heterogeneous high-mobility environments found in \ac{V2X}, cell-free/cooperative multi-cell, and \ac{NTN} communication scenarios \cite{Wu_Access16, Zhou_IEEE20, Wang_CM22}.

A promising enabling technology to satisfy the demands of \ac{B5G} and \ac{6G} is \ac{ISAC}, which combines sensing and communications functions under a single wireless system, with unified hardware and signal processing techniques \cite{Zhang_CST22, Wild_Access21, Wang_JCS22, PinTan_JCS21}.

Besides the greater support to services such as environment mapping and node localization, which is of fundamental importance to the aforementioned high-mobility scenarios, various enhancements are expected to be brought by \ac{ISAC} technology, including increased efficiency in spectrum, energy, and hardware costs.

% CONTINUE FROM HERE

High-mobility environments are a great challenge to wireless communications systems due to the resulting \textbf{doubly-dispersive wireless channel} \cite{Bliss_13, Vitetta_13}, also referred to as \ac{TV} multipath, or time-frequency selectivity.
Although the delays and Doppler shifts can be estimated \cite{SiebertIRE1956, RichardsBook2005, PatoleSPMag2017}, the effect of such a scattering environment onto received signals is severe \ac{ICI} \cite{Wang_TWC06, Matz_06}, which can drastically decrease the communication performance under conventional of even highly effective modulation schemes such as \ac{OFDM}.
And despite various clever contributions that have been proposed to mitigate the issue \cite{Lu_TVT18, Ma_CL20, Keskin_JSTSP21}, hefty consequences in overhead and increased complexity are inevitable.

Motivated by the above challenge, as well as the potential of \ac{ISAC} technology within \ac{B5G} and \ac{6G} systems, novel waveforms have recently been proposed which are inherently robust to high-mobility environments, thanks to the orthogonality they maintain in the doubly-dispersive channel.
One of the best-known and investigated methods with such features is the \ac{OTFS} waveform \cite{Hadani_WCNC17, Wei_WC21}, which leverages the \ac{ISFFT} \cite{Grac_92} to modulate a \ac{2D} grid of information symbols directly onto the delay-Doppler domain.
The \ac{OTFS} scheme gained great attention as a promising candidate for high-mobility \ac{B5G} systems, thanks to the demonstrated superior performance it achieves in comparison to other proposed waveforms \cite{Murali_ITA18, Anwar_WCNC20, Wiffen_PIMRC18}.

Alternatively, chirp domain-based multicarrier waveforms have also been investigated, which are attractive due to their inherent spread-spectrum property and potential for full-duplex operations \cite{Klauder_BST60, Capus_IET00, Wang_JCS22}.
This effort led to the proposal of several novel modulation schemes for doubly-dispersive wireless channel \cite{Ouyang_TC16, Erseghe_TC05, Martone_TV01}, including the recently proposed \ac{AFDM} waveform \cite{Bemani_ICC21, Bemani_TWC23}, which leverages the \ac{IDAFT} \cite{Healy_LCT15} to modulate information symbols into a ``twisted'' time-frequency domain in order to achieve the desired delay-Doppler orthogonality.
This particular feature of \ac{AFDM}, as well as other properties such as the full-diversity guarantee, optimizable parametrization, increased throughput, and reduced computational dimension \cite{Bemani_TWC23}, promotes the \ac{AFDM} approach as a strong contender to \ac{OTFS}.

It was quickly noticed that the full delay-Doppler representation of the channel in \ac{OTFS} and \ac{AFDM} \textbf{inherently conveys the velocity and range information of the scatterers} in the form of the respective multipath delays and Doppler-shifts, such to imply significant benefits in terms of \ac{ISAC}.
This was followed by a plethora of \ac{OTFS}-based \ac{ISAC} techniques being proposed to extract the delay and Doppler parameters of the resolvable paths directly from the \ac{CSI} \cite{Raviteja_Radarconf19, Gaudio_Radarconf20, Keskin_ICC21}, which have been shown to approach the sensing performances of the full \ac{OFDM} and \ac{FMCW} radars, with a higher robustness to mobility and achievable capacity \cite{Gaudio_TWC20}.

Naturally, the \ac{AFDM} waveform is also expected to be a very promising candidate for \ac{ISAC} in doubly-dispersive environments \cite{Wang_JCS22} with similarities and advantages over \ac{OTFS}, in addition to the natural relationship between chirp waveforms and radar signal processing.
However, very few works have been published so far on \ac{AFDM}-based \ac{ISAC} \cite{Ni_ISWCS22}, which is likely due to the fact that the \ac{AFDM} technique was only very recently proposed.
In view of all the above, the full article here described aims to offer a thorough analysis on the future of \ac{ISAC} in heterogeneous high-mobility environments, in the form of a comprehensive comparison of \ac{AFDM} as a rising competitor, and \ac{OTFS} as its leading \ac{SotA} alternative. 

%%%%%%%%%%
\newpage 
\textbf{The proposed contributions of the full article are}:

\begin{itemize}
\item An introduction to \ac{ISAC}, especially in high-mobility environments, highlighting the motivation, challenges, and implications to the consequent signal processing.
\item A \ac{SotA} survey of the novel waveforms designed for high-performance communication in doubly-dispersive channels, exemplified by the \ac{AFDM} and \ac{OTFS} waveforms.
\item A thorough comparison and analysis of the \ac{SotA} \ac{ISAC} techniques leveraging the novel waveforms and other relevant methods such as \ac{FMCW} radar and \ac{OFDM}, highlighting various factors such as computational complexity, radar resolution, spectral efficiency, pilot overhead, etc.
\item A conclusive summary with the potential research directions and remaining challenges of \ac{ISAC} in high-mobility environments with doubly-dispersive channels.
\end{itemize}

\vspace{1ex}
%%%%%%%%%%%%%%%%%%%%%%%%%%%%%%%%%%%%%%%%%%%%%%%%%%%%%%%%%%%%%%%%%
%%%%%%%%%%%%%%%%%%%%%%%%%%%%%%%%%%%%%%%%%%%%%%%%%%%%%%%%%%%%%%%%%
\section{\textbf{Outline of the proposed Special Issue paper}}
\label{secpr}

\vspace{1ex}
\subsection{\textbf{Background on Communications over Doubly-Dispersive (TV Multipath) Channels}}

Consider a wireless channel in a high-mobility scattering environment with $P$ resolvable propagation paths, where each $p$-th path with $p \in \{1,\cdots\!,P\}$ is respectively described by a corresponding channel fading coefficient $h_p \in \mathbb{C}$, delay $\tau_p \in [0, \tau^\mathrm{max}]$, and Doppler shift $\nu_p \in [-\nu^\mathrm{max}, +\nu^\mathrm{max}]$.
The consequent \ac{CIR} of a \ac{TV} multipath channel is typically described by a \textit{time-variant impulse response} function in the time-delay domain,
\begin{equation}
g(t,\tau) \triangleq \textstyle\sum_{p=1}^{P} h_p \!\cdot\! e^{j2\pi \nu_p t} \!\cdot\! \delta(\tau - \tau_p),
\label{eq:cir_td}
\end{equation}  
where $t$ and $\tau$ respectively denote the instantaneous time and delay, and $\delta(x)$ is the unit impulse function.

Alternatively, the \ac{CIR} $g(t,\tau)$ in the time-delay domain can be represented in the dual domains \cite[Ch.2]{Vitetta_13} by leveraging the \ac{FT} $\mathcal{F} \{ f(x) \}$.
For example, the \ac{CIR} can be represented as the \textit{time-variant transfer function} $g^{\mathrm{TF}}(t,f)$ or the \textit{delay-Doppler spread function} $g^{\mathrm{DD}}(\tau,\nu)$, by performing a \ac{FT} over the delay domain or the time domain, respectively, \textit{i.e.,}
\noindent\begin{minipage}[H]{0.47\textwidth}
\begin{equation}    
\!\!\!g^{\mathrm{TF}}(t,f) \!\triangleq \!\! \underset{\tau \rightarrow f}{\mathcal{F}}\! \big\{ g(t,\tau)\big\} \!=\! \textstyle\sum_{p=1}^{P} h_p \!\!\!\cdot\! e^{j2\pi \nu_p t} \!\cdot\!  e^{-j2\pi \tau_p f}\!,
\label{eq:cir_tf} 
\end{equation}
\end{minipage}
\hfill
\begin{minipage}[H]{0.52\textwidth}
\begin{equation}
~g^{\mathrm{DD}}(\tau, \nu) \! \triangleq \!\!\!\underset{t \rightarrow \nu}{\mathcal{F}}\! \big\{ g(t,\tau)\big\} \;\!\! =\!\! \textstyle\sum_{p=1}^{P} \!h_p \!\cdot\! \delta(\tau - \tau_p) \cdot \delta(\nu - \nu_p),
\label{eq:cir_dd}
\end{equation}
\end{minipage}
\vspace{0.05ex}

\noindent where $f$ and $\nu$ respectively denote the instantaneous frequency and Doppler shift.

The two representations of the doubly-dispersive \ac{CIR} given by eq. \eqref{eq:cir_tf} and eq. \eqref{eq:cir_dd} are respectively visualized in Fig. \ref{fig:cir_reps}, highlight the continuously fast-varying nature in the time-frequency domain, in contrast to the sparse and impulsive property in the delay-Doppler domain, with the implication of an inherently more efficient signal processing in the latter.

Following the above, given a discrete sequence of the transmit signal $s[n]$ sampled at a rate of $T$ with discrete time indices $n \in \{0, \cdots\!, N - 1\}$, the received signal over the \ac{CIR} in eq. \eqref{eq:cir_td} is given by
\begin{equation}
r[n] = {\textstyle\sum_{\ell = 0}^{\infty}} \big( s[n - \ell] \cdot \textstyle\sum_{p = 1}^{P} h_p \cdot e^{j2\pi \nu_p \!\frac{n}{N}} \cdot \delta(\tfrac{\tau_p}{\Delta\tau} - \ell)  \big) + w[n],
\label{eq:rec_signal}
\end{equation}
where $w[n]$ is the \ac{AWGN} signal, and $\ell \triangleq \!\!\frac{\tau}{\Delta \tau}$ is the normalized delay with resolution $\Delta \tau$ assumed to be sufficient such the normalized delays can be rounded to the nearest integer with negligible error.

Omitting the full derivations at this point, eq. \eqref{eq:rec_signal} can also be described with as circular convolution by
\begin{equation}
\mathbf{r} \triangleq \mathbf{H}\mathbf{s} + \mathbf{w} = \big( \textstyle\sum_{p=1}^{P} h_p \!\cdot\! \mathbf{C}_{p} \!\cdot\! \mathbf{N}_p \!\cdot\! \mathbf{\Pi}^{\ell_p} \big) \!\cdot \mathbf{s} + \mathbf{w} \in \mathbb{C}^{N \times 1},
\label{eq:rec_vector}
\end{equation}
where $\mathbf{r} \!\in\! \mathbb{C}^{N \times 1}$ is the received signal vector; and $\mathbf{C}_{p} \!\in\! \mathbb{C}^{N \times N}$ is a diagonal matrix arising from a cyclic prefix, $\mathbf{N}_{p} \!\in\! \mathbb{C}^{N \times N}$ is a diagonal Doppler shift matrix, and $\mathbf{\Pi} \in \mathbb{C}^{N \times N}$ is the forward cyclic-shift matrix, respectively for the $p$-th path.

In other words, the resultant delay-Doppler representation of the doubly-dispersive channel is a superposition of $P$ different shifted diagonal matrices.

\vspace{-4ex}
\noindent\begin{minipage}[H]{0.48\textwidth}
\begin{figure}[H]
\centering
\subfloat[Time-frequency transfer function.]{\includegraphics[width=0.49\textwidth]{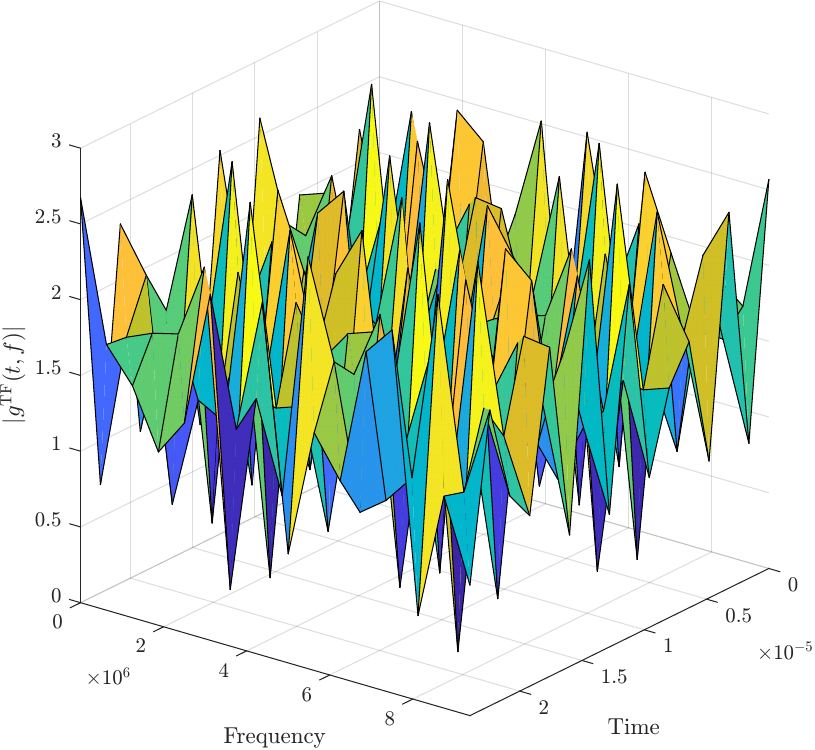}%
\label{fig:cir_tf}}
\hfil
\subfloat[Delay-doppler spread function.]{\includegraphics[width=0.49\textwidth]{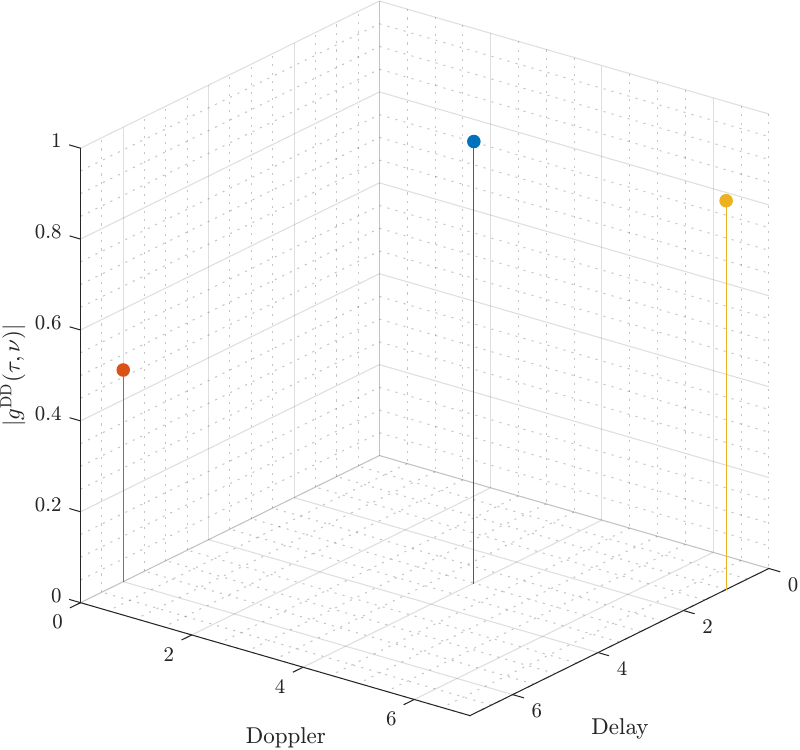}%
\label{fig:cir_dd}}
\caption{Different representations of a doubly-dispersive \ac{CIR} with $P = 3$ resolvable paths (illustrated by unique colors) of random integer delays and Doppler-shifts, with carrier frequency of $5.9$GHz and signal bandwidth of $10$MHz (following the IEEE 802.11p vehicular environment \cite{IEEE802_11p_standard}).}
\label{fig:cir_reps}
\end{figure}
\end{minipage}
\hfil
\begin{minipage}[H]{0.48\textwidth}
\begin{figure}[H]
\centering
\subfloat[\ac{OTFS} effective channel.]{\includegraphics[width=0.49\textwidth]{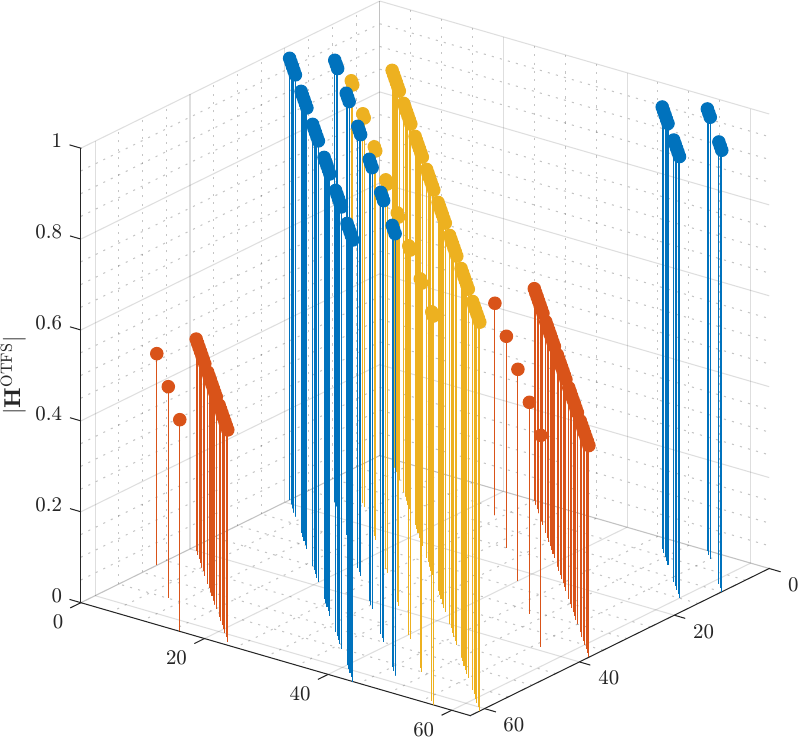}%
\label{fig:cir_otfs}}
\hfil
\subfloat[\ac{AFDM} effective channel.]{\includegraphics[width=0.49\textwidth]{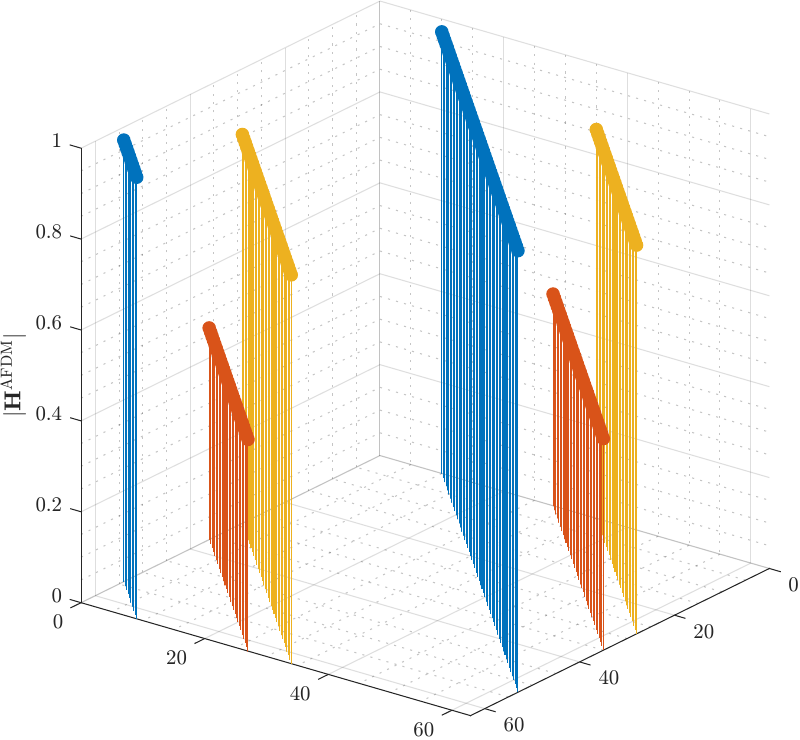}%
\label{fig:cir_ofdm}}
\caption{Illustration of the \ac{OTFS} and \ac{AFDM} effective channels of the exemplary \ac{CIR} in Fig. \ref{fig:cir_reps}, which is a simple case with integer multiple delays and Doppler shifts.
In case of fractional Doppler shifts, the diagonals of the effective matrices would become a thicker band, retaining other properties.}
\label{fig:eff_channel_viz}
\end{figure}
\end{minipage}

%%%%%%%%%%%%%%%%%%%%%%%%%%%%%%%%%%%%%%%%%%%%%%%%%%%%%%%%%%
\subsection{\textbf{Signal Models of OTFS and AFDM Waveforms}}
\subsubsection{Orthogonal Time Frequency Space (OTFS) Modulation} $~$ \vspace{0.5ex}

In \ac{OTFS}, the information symbols are first placed directly on a $K \times L$ grid in the delay-Doppler domain, which is first transformed into the time-frequency domain via the \ac{ISFFT} \cite{Grac_92}, then pulse-shaped into the continuous time signal via the \ac{HT} \cite{Mecklenbruker_ATO89}.
%\footnote{The two-step \ac{OTFS} modulation process can also be implemented as a single \ac{IDZT} \cite{Janssen_88, Heil_89}.}.
%
Mathematically, the \ac{OTFS} modulation process described by
\begin{align}
\mathbf{s}^{\mathrm{OTFS}} \triangleq  \mathrm{vec} \big( \hspace{-2ex} \overbrace{\mathbf{G}^{\mathrm{tx}} \mathbf{F}\herm_{\!K}}^{\text{Pulse-shaped HT}} \hspace{-2ex} \cdot \overbrace{\mathbf{F}^{}_{\!K} \mathbf{X} \mathbf{F}_{\!L}\herm }^{\text{ISFFT}} \big) = (\mathbf{F}\herm_{\!L} \otimes \mathbf{G}^{\mathrm{tx}}) \cdot \overbrace{\mathrm{vec}(\mathbf{X})}^{\triangleq \mathbf{x}} \in \mathbb{C}^{KL \times 1},
\label{eq:tx_otfs}
\end{align} 
where $\mathbf{X} \in \mathbb{C}^{K \times L}$ is the information symbol matrix, $\mathbf{G}^{\mathrm{tx}} \in \mathbf{C}^{K \times K}$ is the diagonal transmit pulse-shaping filter matrix, and $\mathbf{F}_{\!K} \in \mathbb{C}^{K \times K}$ and $\mathbf{F}_{\!L} \in \mathbb{C}^{L \times L}$ are respectively the $K$-point and $L$-point normalized \ac{DFT} matrix.

Given $KL = N$, the filtered and demodulated signal $\mathbf{y}^{\mathrm{OTFS}}$ after the convolution channel in eq. \eqref{eq:rec_signal} is given by
\begin{equation}
\mathbf{y}^{\mathrm{OTFS}} \triangleq (\mathbf{F}_L \otimes \mathbf{G}^{\mathrm{rx}}) (\mathbf{H} \mathbf{s}^{\mathrm{OTFS}} + \mathbf{w}) = \mathbf{H}^{\mathrm{OTFS}}\mathbf{x} + \tilde{\mathbf{w}} \in \mathbb{C}^{N \times 1},
\label{eq:rx_otfs}
\end{equation}
where $\mathbf{G}^{\mathrm{rx}} \in \mathbb{C}^{K \times K}$ is the diagonal matched filter matrix, and $\tilde{\mathbf{w}} \in \mathbb{C}^{N \times 1}$ is the \ac{AWGN} after a unitary transform.

The effective \ac{OTFS} channel on the vectorized information symbol matrix $\mathbf{x} \in \mathbb{C}^{N \times 1}$ is therefore given by
\begin{equation}
\mathbf{H}^{\mathrm{OTFS}} \triangleq \textstyle\sum_{p=1}^{P} \big(  (\mathbf{F}_L \otimes \mathbf{G}^{\mathrm{rx}}) \cdot h_p \mathbf{C}_{p} \mathbf{N}_{p} \mathbf{\Pi}^{\ell_p} \cdot (\mathbf{F}\herm_{L} \otimes \mathbf{G}^{\mathrm{tx}})\big) \in \mathbb{C}^{N \times N}
\label{eq:eff_otfs}
\end{equation}
which is a block-wise pulse-shaped \acs{DFT} and \acs{IDFT}, respectively on the rows and columns of the channel from eq. \eqref{eq:rec_vector}.

%%%%%%%%%%%%%%%%%%%%%%%%%%%%%%%%%%%%%%%%%%%%%%%%%%%%%%%%%%%%%%%%%
\vspace{1ex}
\subsubsection{Affine Frequency Division Multiplexing (AFDM)}$~$ \vspace{0.5ex}

On the other hand, the \ac{AFDM} unlike the \ac{OTFS}, directly multiplexes a one-dimensional vector of symbols unto a \textit{twisted} time-frequency chirp domain using the \ac{IDAFT} \cite{Pei_TSP00}, which is described by
\begin{equation}
\mathbf{s}^{\mathrm{AFDM}} \triangleq \overbrace{(\mathbf{\Lambda}_{c_1}\herm \mathbf{F}_N\herm \mathbf{\Lambda}_{c_2}\herm )}^{\text{IDAFT}} \!\;\!\mathbf{x} = (\mathbf{\Lambda}_{c_2} \mathbf{F}_N \mathbf{\Lambda}_{c_1} \!)^{-\!1} \mathbf{x} = \mathbf{A}^{\!-1} \mathbf{x} \in \mathbb{C}^{N \times 1},
\label{eq:tx_afdm}
\end{equation} 
where $\mathbf{A} \triangleq \mathbf{\Lambda}_{c_2} \mathbf{F}_N \mathbf{\Lambda}_{c_1} \in \mathbb{C}^{N \times N}$ is the forward $N$-point \ac{DAFT} matrix with the $N$-point normalized \ac{DFT} matrix $\mathbf{F}_{N} \in \mathbb{C}^{N \times N}$, and $\mathbf{\Lambda}_{c_i} \triangleq \mathrm{diag}[e^{-j2\pi c_i (0)^2}, \cdots, e^{-j2\pi c_i (N-1)^2}] \in \mathbb{C}^{N \times N}$ is a diagonal discrete quadratic-chirp matrix of central digital frequency $c_i$.

Consequently, the demodulated \ac{AFDM} signal over the convolution channel in eq. \eqref{eq:rec_vector} is given by
\begin{equation}
\mathbf{y}^{\mathrm{AFDM}} = \mathbf{A} \big( \textstyle\sum_{p=1}^{P} h_p  \mathbf{C}_{p} \mathbf{N}_p \mathbf{\Pi}^{\ell_p} \big)\! \cdot \mathbf{s}^{\mathrm{AFDM}}+ \tilde{\mathbf{w}} = \mathbf{H}^{\mathrm{AFDM}}\mathbf{x} + \tilde{\mathbf{w}} \in \mathbb{C}^{N \times 1},
\label{eq:rx_afdm}
\end{equation} 
where the effective \ac{AFDM} channel on the information symbol vector $\mathbf{x} \in \mathbb{C}^{N \times 1}$ correspondingly defined as
\begin{equation}
\mathbf{H}^{\mathrm{AFDM}} \triangleq \textstyle\sum_{p=1}^{P} \! \big( \mathbf{A} (  h_p \mathbf{C}_{p} \mathbf{N}_p \mathbf{\Pi}^{\ell_p} ) \mathbf{A}^{\!-1} \big) \in \mathbb{C}^{N \times N}.
\label{eq:eff_afdm}
\end{equation} 

The consequent effect of the \acs{DAFT}/\acs{IDAFT} and the block-wise \acs{DFT}/\acs{IDFT} unto the channel in eq. \eqref{eq:rec_vector} can be observed in Fig. \ref{fig:eff_channel_viz}, where the simple exemplary effective channels of the \ac{OTFS} and \ac{AFDM} schemes have been illustrated.

\subsection{\textbf{Radar Parameter Estimation via \ac{AFDM} and \ac{OTFS} Waveforms }}
\label{sec:radar_estimation}

In light of the above, to enable \ac{ISAC} through the aforementioned waveforms, the extraction of the radar parameters of the scattering targets from the inherent channel information is necessary.

As one example in this white paper, we provide the widely-utilized \ac{SotA} method estimate the channel parameters, which aims to solve the \ac{ML} problem on the channel parameters given by
\begin{equation}
\underset{\boldsymbol{\theta}}{\mathrm{arg ~min}} \;\big\| \mathbf{y} - \textstyle\sum_{p=1}^{P} \tilde{\mathbf{H}}(h_p, \tau_p, \nu_p) \!\cdot\! \mathbf{x} \big\|^2_2,
\label{eq:ML}
\end{equation}
where $\boldsymbol{\theta} = \{h_1,\cdots\!,h_P,\tau_1, \cdots\!,\tau_P, \nu_1, \cdots\!,\nu_P\}$ is the set of all $3P$ channel parameters corresponding to the channel gain, delay, and Doppler shift of the $P$ paths, respectively, and $\tilde{\mathbf{H}}(h_p, \tau_p, \nu_p)$ is the conditional channel matrix given the parameters $h_p, \tau_p, \nu_p$, defined as the channel in eq. \eqref{eq:rec_vector}.

The joint domain of $3P$ parameters, which are continuously complex on $h_p$ and bounded but still continuously real on $\tau_p$ and $nu_p$, for all $P$, the above minimization problem is extremely challenging.
One efficient widely utilized framework is the iterative grid search or reduced-search methods \cite{Gaudio_Radarconf20,Tang_Access23} which are known to exhibit a trade-off in complexity and resolution.
On the other hand, other methods have also investigated leveraged subspace analysis and radar-like signal processing methods \cite{Qu_TC21, Liu_TC22, Zacharia_PC23}  to directly approximate the continuous parameters of eq. \eqref{eq:ML} from the effective channel information.

The various approaches and the advantages of the available methods will be discussed in the full article - specifically aiming to provide relevant results and evaluation for the \ac{AFDM}-based sensing which is currently lacking in the relevant literature.

%%%%%%%%%%%%%%%%%%%%%%%%%%%%%%%%%%%%%%%%%%%%%%%%%%%%%%%%%
\subsection{\textbf{Comparative Analysis of \ac{OTFS} and \ac{AFDM} Schemes}}

Finally, in light of the above sections including the signal models and \ac{ISAC} methods of the two waveforms, few differences in the inherent signal processing between the two schemes can already be highlighted, such as the nature of the underlying transforms and its effect on the effective channel.
Such differences between the \ac{OTFS} and the \ac{AFDM} waveforms are briefly provided in the below Table \ref{tab:comp}, where a few other properties which have not been described in this overview white paper also provided, including modulation complexity, asymptotic transmit diversity, and required pilot overhead.

We emphasize that the short comparative table below is only an anticipation of the thorough investigation to be expected in the full article, which will additionaly compare sensing capacity, total throughput, channel/data estimation methods, \emph{etc.}, and also include other relevant methods such as the \ac{OFDM} and \ac{OCDM}.

In hand of the final comparative analysis, a conclusive summary and the identification of the remaining challenges and future directions of research will be provided, to constructively close the full loop of the proposed article.

\begin{table}[H]
\vspace{-2.5ex}
\caption{\ul{Brief} comparison of waveform properties of \ac{AFDM} and \ac{OTFS}.}
\label{tab:comp}
\vspace{-2.5ex}
\centering
\begin{tabular}{|c|c|c|}
\hline \hline
Property & AFDM & OTFS \\ \hline\hline
Transform Domain & Multicarrier chirp & Delay-Doppler \\ \hline
Eff. Channel Structure & Shifted band & Scattered diagonal \\  \hline
Modulation Complexity & $N\log_2\!N + 2N$ & $(3/2)N\log_2\!N$\\ \hline
Asymptotic Diversity & Full diversity \cite{Bemani_TWC23} & Order-$1$ \cite{Surabhhi_TWC19} \\  \hline
Pilot Guard Overhead & 1D zero-pad & 2D zero-pad \\ \hline
\hline
\end{tabular}
\end{table}
\vspace{-1ex}

%%%%%%%%%%%%%%%%%%%%%%%%%%%%%%%%%%%%%%%%%%%%%%%%%%%%%%%%%%%%%%%%%
%%%%%%%%%%%%%%%%%%%%%%%%%%%%%%%%%%%%%%%%%%%%%%%%%%%%%%%%%%%%%%%%%
%\bibliographystyle{IEEEtran}
%\bibliography{listrefs_HSRou}

%%%%%%%%%%%%%%%%%%%%%%%%%%%%%%%%%%%%%%%%%%%%%%%%%%%%%%%%
%%%%%%%%%%%%%%%%%%%%%%%%%%%%%%%%%%%%%%%%%%%%%%%%%%%%%%%%%%%%%%%%%%%%%%%%%%%%%
\end{document}